\documentclass[a4paper,fleqn]{cas-dc}

\usepackage{hyperref}
\usepackage{graphicx}
\usepackage[caption=false]{subfig}
\usepackage{lipsum}
\usepackage{lscape}
\usepackage{amsmath}
\usepackage{amssymb}
\usepackage{multirow}
\usepackage[figuresright]{rotating}
\usepackage{algorithm} 
\usepackage{algpseudocode} 
\usepackage{comment}
\usepackage{lineno}
\usepackage{nomencl}
\makenomenclature
\usepackage{calc}
\usepackage[T1]{fontenc}
\usepackage{relsize}
\usepackage{orcidlink}
\usepackage{bm}

\usepackage[sort,comma,authoryear,round]{natbib}

\setcitestyle{authoryear,open={(},close={)},citesep={;}} 

\hypersetup{
  colorlinks,
  citecolor=Violet,
  linkcolor=Red,
  urlcolor=Blue}

\newcolumntype{P}[1]{>{\centering\arraybackslash}p{#1}}

\def\tsc#1{\csdef{#1}{\textsc{\lowercase{#1}}\xspace}}
\tsc{WGM}
\tsc{QE}
\tsc{EP}
\tsc{PMS}
\tsc{BEC}
\tsc{DE}

\begin{document}
\let\WriteBookmarks\relax
\def\floatpagepagefraction{1}
\def\textpagefraction{.001}
\shorttitle{3D Reconstruction in Plant Phenotyping}
\shortauthors{Jiajia Li, etc. (2025)}
 
\title [mode = title]{A Survey on 3D Reconstruction Techniques in Plant Phenotyping: From Classical Methods to Neural Radiance Fields (NeRF), 3D Gaussian Splatting (3DGS), and Beyond}

\author[1]{Jiajia Li\orcidlink{xx}}\ead{lijiajia@msu.edu}
\author[1]{Xinda Qi\orcidlink{xx}}\ead{qixinda@msu.edu}
\author[2]{Seyed Hamidreza Nabaei\orcidlink{XX}}\ead{fgx9eq@virginia.edu}
\author[3]{Meiqi Liu\orcidlink{XX}}\ead{liumeiqi@msu.edu}
\author[4]{Dong Chen\orcidlink{0000-0002-0551-5084}}\ead{dc2528@msstate.edu}
\author[4]{Xin Zhang\orcidlink{XX}}\ead{xzhang@abe.msstate.edu}
\author[5]{Xunyuan Yin\orcidlink{XX}}\ead{xunyuan.yin@ntu.edu.sg}
\author[6]{Zhaojian Li*\orcidlink{XX}}\ead{lizhaoj1@egr.msu.edu}

\address[1]{Department of Electrical and Computer Engineering, Michigan State University, East Lansing, MI, USA}
\address[2]{Department of Systems and Information Engineering \& Link Lab, University of Virginia, Charlottesville, VA, USA}
\address[3]{Department of Statistics and Probability, Michigan State University, East Lansing, MI, USA}
\address[4]{School of Chemistry, Chemical Engineering and Biotechnology, Nanyang Technological University, Singapore}
\address[5]{Department of Agricultural and Biological Engineering, Mississippi State University, Starkville, MS, USA}
\address[6]{Department of Mechanical Engineering, Michigan State University, East Lansing, MI, USA}

\address{* Corresponding author}

\begin{abstract}
Plant phenotyping plays a pivotal role in understanding plant traits and their interactions with the environment, making it crucial for advancing precision agriculture and crop improvement. 3D reconstruction technologies have emerged as powerful tools for capturing detailed plant morphology and structure, offering significant potential for accurate and automated phenotyping. This paper provides a comprehensive review of the 3D reconstruction techniques for plant phenotyping, covering classical reconstruction methods, emerging Neural Radiance Fields (NeRF), and the novel 3D Gaussian Splatting (3DGS) approach. Classical methods, which often rely on high-resolution sensors, are widely adopted due to their simplicity and flexibility in representing plant structures. However, they face challenges such as data density, noise, and scalability.
NeRF, a recent advancement, enables high-quality, photorealistic 3D reconstructions from sparse viewpoints, but its computational cost and applicability in outdoor environments remain areas of active research. The emerging 3DGS technique introduces a new paradigm in reconstructing plant structures by representing geometry through Gaussian primitives, offering potential benefits in both efficiency and scalability. We review the methodologies, applications, and performance of these approaches in plant phenotyping and discuss their respective strengths, limitations, and future prospects (\url{https://github.com/JiajiaLi04/3D-Reconstruction-Plants}). Through this review, we aim to provide insights into how these diverse 3D reconstruction techniques can be effectively leveraged for automated and high-throughput plant phenotyping, contributing to the next generation of agricultural technology.
\end{abstract}


\maketitle

\begin{table}[]
\centering
\renewcommand{\arraystretch}{1.4}
\centering
\caption{Nomenclature}
\label{tab:nomen1}
\resizebox{0.4\textwidth}{!}{%
\begin{tabular}{|ll|}
\hline
Nomenclature &                                                                     \\ \hline
NeRF         & Neural Radiance Field                                               \\ 
3DGS         & 3D Gaussian Splatting                                               \\ 
MVS          & Multi-View Stereo                                                   \\
SLAM         & Simultaneous Localization and Mapping                               \\
SfM          & Structure-from-Motion                                               \\ 
SIFT         & Scale Invariant Feature Transform                                   \\ 
PMVS         & PatchMatch Multi-View Stere                                         \\ 
PSNR         & Peak Signal-to-Noise Ratio                                          \\ 
SSIM         & Structural Similarity Index Measure                                 \\ 
LPIPS        & Learned Perceptual Image Patch Similarity                           \\ 
IoU          & Intersection over Union                                             \\ 
CD           & Chamfer Distance                                                    \\ 
BO           & Boundary Overlap                                                    \\ 
$R^2$         & Coefficient of Determination                                        \\ 
RMSE         & Root Mean Squared Error                                             \\ 
MAPE         & Mean Absolute Percentage Error                                      \\ 
SAM          & Segment Anything Model \\ 
HSI          & Hyperspectral imaging                                               \\ 
VR           & Virtual Reality                                                     \\ 
AR           & Augmented Reality                                                   \\ \hline
\end{tabular}
}
\end{table}

\section{Introduction}
\label{sec:intro}
Plant phenotype plays an important role in genetics, botany, and agronomy, involving the measurement and analysis of plant's physical and biochemical characteristics \citep{li2020review}. 
This process involves the systematic assessment of various traits, including leaf width and length, plant size, volume, root structure, flowering time, fruit volume, and other features that collectively determine plant growth, development, and overall performance \citep{akhtar2024unlocking}.
Accurate phenotypic analysis is particularly valuable for plant breeders, enabling the efficient identification and selection of desirable traits to enhance crop yield, quality, and stress tolerance \citep{saeed2023peanutnerf}. 

Traditionally, manual phenotypic measurement and analysis are time-consuming, labor-intensive, and costly, limiting the scale and efficiency of phenotyping programs. However, recent advancements in reliable, automatic, multifunctional, and high-throughput phenotyping technologies are transforming the field, enabling rapid and precise trait evaluation across large populations of plants \citep{li2020review}. 
Among them, 2D imaging has been widely adopted using RGB cameras to capture key morphological traits such as size, shape, and color variations at affordable prices. However, these methods are inherently limited in its ability to capture complex three-dimensional plant structures, leading to potential inaccuracies in traits like biomass estimation and canopy architecture reconstruction \citep{hu2024high}.

To overcome the limitations of traditional imaging, 3D imaging technologies, particularly active reconstruction methods, have been developed to capture detailed structural information. These methods rely on depth sensors to generate 3D coordinates by sampling points on the surface of target objects \citep{arshad2024evaluating}. By providing depth data, they enable more accurate and comprehensive phenotypic analysis, enhancing the precision of trait quantification in complex plant architectures.
For instance, RGB-D cameras can simultaneously capture both color (RGB) and depth information, making them particularly useful for greenhouse and indoor phenotyping applications, where controlled lighting conditions improve depth-sensing accuracy \citep{vit2018comparing, choi2024nerf}. LiDAR, on the other hand, measures distances using laser pulses to produce highly accurate 3D point clouds, making it well-suited for large-scale outdoor phenotyping. Its ability to penetrate dense canopies and capture fine structural details has made it an invaluable tool for characterizing tree architecture and crop morphology \citep{lin2015lidar}.
Passive reconstruction methods, such as stereo vision and photogrammetry—also contribute to 3D phenotyping \citep{schonberger2016structure}. However, despite the promise of both active and passive approaches, existing 3D phenotyping methods still face considerable challenges. For example, multi-view stereo (MVS)-based reconstruction involves multiple processing steps, resulting in long computation times and limited robustness in variable conditions \citep{wu2022miniaturized}. Likewise, point cloud-based methods using depth sensors often suffer from noise, occlusions, and incomplete data acquisition, especially in complex field environments \citep{hu2024high}.

In recent years, deep-learning-driven approaches such as Neural Radiance Fields (NeRF) and 3D Gaussian Splatting (3DGS) have emerged as cutting-edge techniques for high-fidelity 3D object reconstruction \citep{gao2022nerf, chen2024survey}. These methods hold immense potential for automated, high-throughput phenotyping by enabling rapid and detailed plant reconstructions.
NeRF leverages neural representations to synthesize photorealistic 3D models from sparse input images, capturing fine structural details that traditional methods struggle to reconstruct \citep{mildenhall2021nerf}. Unlike conventional point cloud-based approaches, NeRF models are self-supervised and can be trained using only multi-view images and camera poses without requiring explicit 3D or depth supervision. This makes them particularly advantageous for complex plant architectures, where traditional depth-sensing techniques may suffer from occlusions or noise. 
Similarly, 3DGS introduces an innovative paradigm by representing plant structures with Gaussian primitives, offering significant advantages in rendering speed and scalability \citep{kerbl20233d}. Unlike NeRF, which relies on volumetric rendering, 3DGS achieves efficient real-time reconstruction and visualization, making it more suitable for high-throughput phenotyping applications.
Despite the promising advancements, the application of NeRF and 3DGS in plant phenotyping remains an emerging research area, gaining increasing attention from the agriculture community. 

Existing review papers on plant phenotyping primarily focus on well-established 3D imaging techniques, such as LiDAR, structured light, and multi-view stereo (MVS). For instance, \cite{fiorani2013future, li2014review} provide comprehensive reviews of these conventional imaging techniques and their applications in plant phenotyping, offering valuable insights into their strengths and limitations.  Additionally, \cite{li2020review, jiang2020convolutional} provide overviews of the typical computer vision technologies and datasets used in the field. Moreover, \cite{costa2019plant} analyzes the research trends in plant phenotyping over the past two decades, including trends in authorship and the geographical distribution of research activity across countries.
While these reviews contribute valuable perspectives on traditional methods and the broader trends in the field, they do not sufficiently address the growing role of deep learning approaches, such as NeRF and 3DGS, which have the potential to revolutionize plant phenotyping by enabling more accurate, efficient, and scalable solutions. 
Table~\ref{tab:reviews} provides a summary of existing literature reviews on plant phenotyping.

This paper, therefore, aims to provide a systematic review of the emerging role of deep-learning-driven 3D reconstruction methods, such as NeRF and 3DGS, in plant phenotyping. We survey recent developments in these 3D reconstruction techniques and their applications in plant phenotyping, highlighting the technical challenges and future research needs. This review will be beneficial for the research community, offering valuable insights into the potential of these cutting-edge methods to advance plant phenotyping and guide future investigations in this field.

The rest of the paper is organized as follows. Section~\ref{sec2} reviewed the preliminaries of recent 3D reconstruction techniques, while Section~\ref{sec3} introduces key applications of these methods in plant phenotyping. Section~\ref{sec4} discusses the challenges and future directions, with a concluding summary provided in Section~\ref{sec5}. In Appendix, we detail the method for methodically and exhaustively compiling related studies.

\begin{table}[]
\centering
\renewcommand{\arraystretch}{1.3}
\caption{Summary of existing literature reviews on plant phenotyping.}
\label{tab:reviews}
\resizebox{0.48\textwidth}{!}{%
\begin{tabular}{|l|c|c|c|c|}
\hline
                          & \multicolumn{2}{c|}{Classical methods}                               &                                  &                                  \\ \cline{2-3} 
\multirow{-2}{*}{Reviews} & Active methods                   & Passive methods                  & \multirow{-2}{*}{NeRF}                              & \multicolumn{1}{c|}{\multirow{-2}{*}{3DGS}}                             \\ \hline \hline
\cite{fiorani2013future}       & \checkmark & - & -                                & -                                \\
\cite{li2014review}            & \checkmark & - & -                                & -                                \\
\cite{costa2019plant}          & - & - & -                                & -                                \\
\cite{das2019leveraging}       & \checkmark & - & -                                & -                                \\
\cite{li2020review}            & \checkmark & \checkmark & -                                & -                                \\
\cite{jiang2020convolutional}  & \checkmark & \checkmark & -                                & -                                \\
\cite{atefi2021robotic}        & \checkmark & - & -                                & -                                \\
\cite{guo2021uas}              & \checkmark & \checkmark & -                                & -                                \\
\cite{saric2022applications}   & - & - & -                                & -                                \\
\cite{kolhar2023plant}         & \checkmark & \checkmark & -                                & -                                \\
\cite{akhtar2024unlocking}     & \checkmark & \checkmark & \checkmark                                & -                                \\
\cite{song2025comprehensive}   & \checkmark & \checkmark & -                                & -                               \\ \hline
Ours                      & \checkmark & \checkmark & \checkmark & \checkmark \\ \hline
\end{tabular}%
}
\end{table}

\section{Preliminaries of 3D Reconstruction}
\label{sec2}
This section presents the preliminaries of 3D reconstruction and the metrics used to evaluate reconstruction performance.

\subsection{Classical methods} \label{sec:2.1}
Classical 3D reconstruction methods can be broadly categorized into \textit{active} and \textit{passive} approaches, both of which are applicable to plant phenotyping.

\subsubsection{Active reconstruction methods}
\label{sec:2.1.1}
\begin{figure*}
    \centering
    \includegraphics[width=0.95\linewidth]{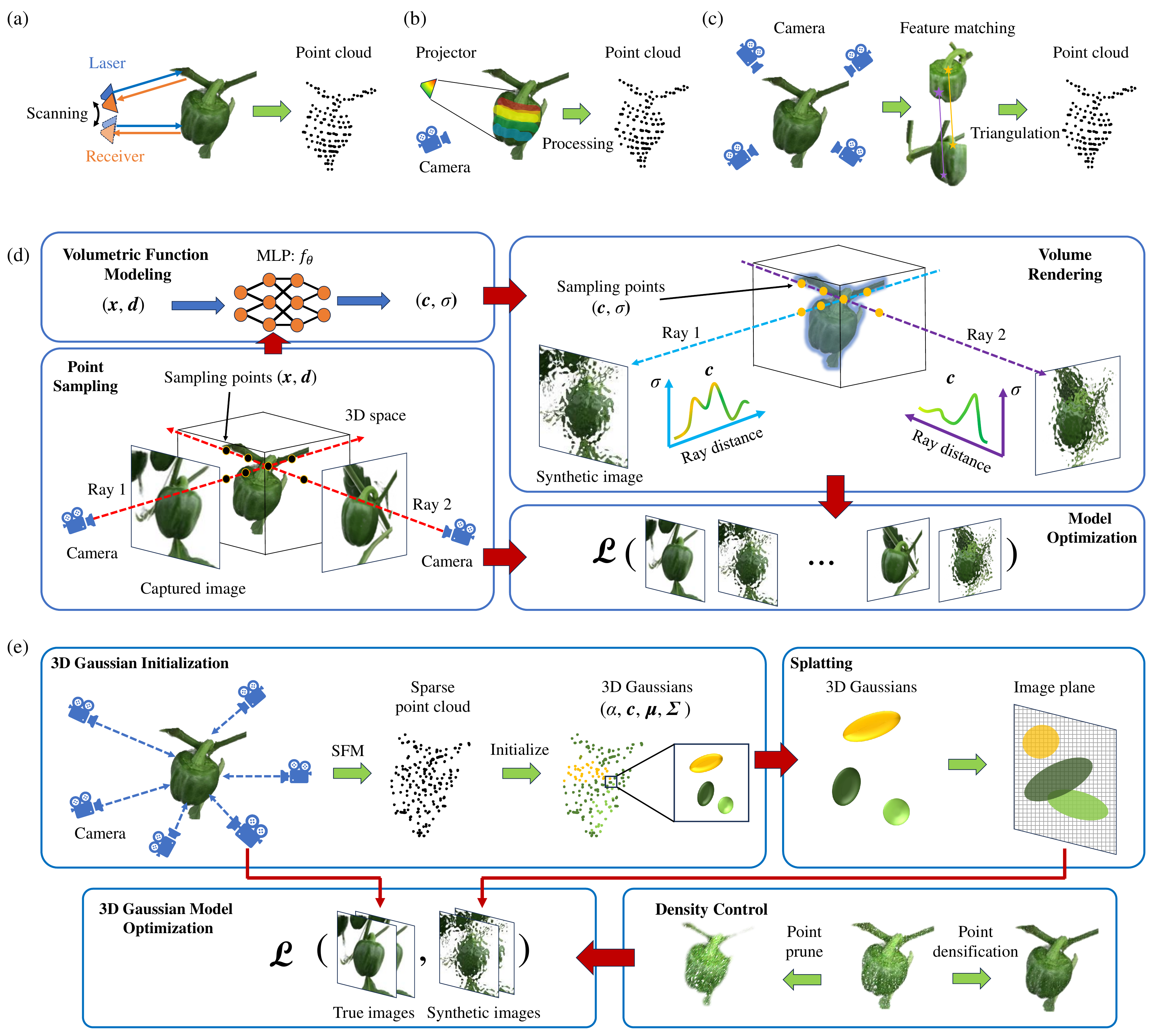}
    \caption{Comparison of different 3D reconstruction techniques and their corresponding reconstruction processes: (a) LiDAR, (b) Structured Light, (c) Structure from Motion (SfM), (d) Neural Radiance Fields (NeRF), and (e) 3D Gaussian Splatting (3DGS).}
    \label{fig:traditional_method}
\end{figure*}

Active reconstruction methods rely on depth sensors to generate 3D coordinates by sampling points on the surface of a target object. Among them, LiDAR \citep{lin2015lidar} is widely used in active methods, such as remote sensing, Simultaneous Localization and Mapping (SLAM), and autonomous systems. LiDAR sensors emit laser pulses in various directions and measure the time interval between emission and reception to compute the depth of each sampled point relative to the sensor, as illustrated in Figure~\ref{fig:traditional_method}(a).  
Various LiDAR technologies have been developed for different application scenarios, including fluorescence LiDAR \citep{svanberg1995fluorescence}, flash LiDAR \citep{rosell2012review}, and high spectral resolution LiDAR \citep{burton2012aerosol}, all of which follow a similar sensing principle. These methods produce sparse 3D point clouds of objects in the scene, which can then be projected onto image planes to synthesize views from different directions.
Despite LiDAR’s advantages, such as high accuracy, precision, and the ability to capture detailed 3D spatial structures crucial for large-scale plant phenotyping, it still faces challenges, including high equipment costs, computational burdens in processing large-scale point cloud data, and limitations in penetrating dense vegetation canopies \citep{tang2024lidar}.

Another widely used active reconstruction technique is 3D structured-light reconstruction \citep{geng2011structured}. In this method, a light projector emits structured light patterns onto the scene, and a camera positioned at a different location, captures images of the scene with the projected patterns, as illustrated in Figure~\ref{fig:traditional_method}(b). The 3D coordinates of the sampling points on the surface of objects in the scene are then computed using principles of geometric optics, based on the captured images and the relative positioning between the projector and the camera.
Similar to LiDAR, structured-light reconstruction generates sparse 3D point clouds of the scene. These point clouds can be further processed into voxel-based or mesh-based 3D models using techniques such as point cloud stitching and surface reconstruction.

\subsubsection{Passive reconstruction methods}
\label{sec:2.1.2}
In addition to active 3D reconstruction methods, which utilize dedicated light sources to illuminate the scene, passive methods can also be employed. These approaches rely solely on image inputs from cameras without the need for external illumination. One representative technique is incremental Structure from Motion (SfM) \citep{schonberger2016structure}, which reconstructs sparse point cloud data from multiple images of the same scene captured from different viewpoints.
As illustrated in Figure~\ref{fig:traditional_method}(c), incremental SfM follows a sequential processing pipeline with an iterative reconstruction approach. It begins by extracting and matching features (e.g., SIFT \citep{lowe2004distinctive}) from overlapping images, then incrementally estimates camera poses and triangulates 3D points to build and expand the model. This process continues by registering additional images and refining both camera parameters and point locations through bundle adjustment until the entire scene is reconstructed.

Other multi-view stereo (MVS) algorithms, such as PatchMatch Multi-View Stereo (PMVS) \citep{romanoni2019tapa}, can further enhance 3D reconstruction quality. These methods refine the initial sparse point cloud generated by SfM by performing dense matching across multiple views to produce denser point clouds or surface grids. In addition, passive techniques like Photometric Stereo \citep{basri2007photometric} are also employed, particularly for reconstructing smooth surfaces. This method estimates surface normals by analyzing the variation in reflected light intensity across images captured under different lighting conditions. 

However, these classical method face challenges in accurately reconstructing scenarios with complex lighting, reflections, and texture less objects, due to the sparse nature of the point clouds and their reliance on 2D features.

\subsection{Neural radiance fields (NeRF)}  \label{sec:2.2}
NeRF is a novel method for synthesizing photo-realistic views of 3D objects by learning a continuous volumetric scene representation \citep{mildenhall2021nerf}. Instead of explicitly reconstructing a 3D model, NeRF leverages a learned radiance field to generate novel views from arbitrary angles, making it a convenient and effective tool for plant phenotyping.
NeRF represents objects in a scene as a continuous volumetric function, which is approximated by a neural network. This function maps spatial coordinates and viewing directions to color and density values. To train the neural network, multiple images of the same scene captured from different viewpoints are required, enabling the optimization of the underlying scene representation.

The input to the network is the 3D location $\bm{x} = (x, y, z)$ and the 2D viewing direction $(\theta, \phi)$ of a sampling point along the camera ray, which can be computed from the known camera poses.
In practice, the viewing direction is often represented as a 3D unit vector $\bm{d} = (d_x, d_y, d_z)$ in Cartesian coordinates. The output of the model is the emitted color $\bm{c} = (r, g, b)$ and the volume density $\sigma$ at that point. The model is therefore represented as:
\begin{equation}
    (\bm{c}, \sigma) = F_\theta(\bm{x}, \bm{d}),
\end{equation} 
where $F_\theta(\cdot)$ is a neural network with learnable parameters $\theta$.

To ensure consistency across multiple views, the volume density $\sigma$ is modeled as a function of only the spatial location $\bm{x}$, while the color $\bm{c}$ depends on both $\bm{x}$ and the viewing direction $\bm{d}$. A hierarchical neural network structure \citep{mildenhall2021nerf} is adopted to enforce this constraint: the network first predicts $\sigma$ from $\bm{x}$, and then uses $\bm{d}$ (along with learned features) to predict $\bm{c}$.

Once trained, the network can infer $\sigma$ and $\bm{c}$ for novel views. A volume rendering technique \citep{kajiya1984ray} is used to compute the expected color $C(\bm{r})$ of a ray $\bm{r}_t = \bm{o} + t\bm{d}$, where $\bm{o}$ is the camera origin and $\bm{d}$ is the view direction, $t$ is the distance on the ray from the $\bm{o}$. The color is calculated as:
\begin{equation}\label{eq:C(r)}
    C(\bm{r}) = \int_{t_1}^{t_2} T(t) \cdot \sigma(\bm{r}(t)) \cdot \bm{c}(\bm{r}(t),d) \text{d}t,
\end{equation}
where $\sigma(\bm{r}(t))$ and $\bm{c}(\bm{r}(t),d)$ denotes the volume density and the emitted color of the sampling point at $\bm{r}(t)$ along the view direction $\bm{d}$, respectively. The bounds $t_1$ and $t_2$ define the near and far limits of the ray.

The function $T(t)$ represents the accumulated transmittance, or the probability that the ray reaches a point, which has a distance of $t$ with respect to $\bm{o}$, without being occluded:
\begin{equation}
    T(t) = \exp{(-\int_{t_1}^t \sigma(\bm{r}(s)) \text{d}u)}.
\end{equation}

In practice, this integral is approximated numerically to render each pixel in the novel view. The ray is divided into $N$ segments, and points are sampled uniformly along the ray. The estimated color is then given by: 
\begin{equation} \hat{C}(\bm{r}) = \sum_{i=1}^N T_i \alpha_i \bm{c}i, \qquad T_i = \exp{\left(-\sum_{j=1}^{i-1} \sigma_j \delta_j \right)}, \end{equation} 
where $T_i$ is the numerical estimation of the accumulated transmittance, $\delta_i$ is the distance between adjacent sample points, and $\alpha_i = 1 - \exp{(-\sigma_i \delta_i)}$ represents the opacity at point $i$.

In addition to RGB rendering, NeRF also enables estimation of depth information $h(\bm{r})$ from the scene using: 
\begin{equation}
    h(\bm{r}) = \int_{t_1}^{t_2} T(t)\sigma(\bm{r}(t))\cdot t\text{d}t,
\end{equation}
where the distance information $t$ is used to replace the emitted color $\bm{c}$ in Eq.~\ref{eq:C(r)}. Similarly, a numerical approximation of the depth is given by: 
\begin{equation}
    \hat{h}(\bm{r}) = \sum_{i=1}^N T_i \alpha_i t_i.
\end{equation}
This approach allows for modeling the 3D geometry of a scene by reconstructing depth from multiple viewpoints.

A typical loss function used to train the neural network is the mean squared error between predicted and ground truth pixel colors:
\begin{equation}
    \mathcal L = \sum_{\bm{r} \in \mathcal R} ||C(\bm{r})-\hat{C}(\bm{r})||_2^2,
\end{equation}
where $\mathcal R$ is the set of rays in each batch. $\hat{C}(\bm{r})$ and $C(\bm{r})$ are the predicted and ground truth color of ray $\bm{r}$, respectively. 

The training process of NeRF, illustrated in Figure~\ref{fig:traditional_method}(d), follows these steps: 
\begin{enumerate} 
    \item Capture multiple images of the same scene from different viewpoints. Sample 3D points and derive position and viewing direction $(\bm{x}, \bm{d})$. 
    \item Use the neural network to estimate color and density $(\bm{c}, \sigma)$ for each sample point. 
    \item Synthesize pixel color $\hat{C}(\bm{r})$ for novel views with the same camera poses and orientations as the ground truth. 
    \item Optimize the model by minimizing the color error between the synthesized and true images. 
\end{enumerate}


\subsection{3D Gaussian splatting (3DGS)}  \label{sec:2.3}
While NeRF-based methods have demonstrated remarkable success in high-fidelity 3D reconstruction, they are often computationally expensive and require extensive training and rendering times. To address these limitations, 3DGS with explicit 3D geometry modeling \citep{kerbl20233d} has emerged as a promising alternative, offering real-time rendering capabilities, efficient scene representation, and high reconstruction accuracy. 

Unlike implicit radiance field methods such as NeRF, which learn a volumetric scene function mapping spatial location and viewing direction $(\bm{x}, \bm{d})$ to color and density $(\bm{c}, \sigma)$ without explicitly defining 3D geometry, explicit radiance field methods directly associate radiance information with points in a 3D geometry representation. 3DGS is an explicit method that models scenes using learnable 3D Gaussian distributions. The model is trained using supervised learning to minimize the discrepancy between synthetic and ground truth multi-view images. The explicit 3D Gaussian representation significantly enhances learning efficiency and enables real-time rendering of complex, high-resolution scenes.

In 3DGS, the geometry of a scene is modeled as a collection of 3D Gaussians (ellipsoids), each defined by its radiance attributes (opacity $\alpha$, color $\bm{c}$) and location attributes (center $\bm{\mu} = (\mu_x, \mu_y, \mu_z)$ and 3D covariance matrix $\bm{\Sigma}$). Unlike NeRF, the opacity $\alpha$ in 3DGS is an independent learnable parameter and not derived from a density value $\sigma$. The color $\bm{c}$ can be modeled using spherical harmonics to capture view-dependent effects.

To ensure that the covariance matrix $\bm{\Sigma}$ is positive semi-definite and interpretable, it is decomposed as: 
\begin{equation} 
    \bm{\Sigma} = \bm{R}\bm{S}\bm{S}^T\bm{R}^T, 
\end{equation} 
where $\bm{R}$ is a rotation matrix (often parameterized by a quaternion), and $\bm{S}$ is a diagonal scaling matrix.

The rendering process involves splatting, where 3D Gaussians are projected onto a 2D image plane. To enhance efficiency, frustum culling \citep{assarsson2000optimized} is used to discard Gaussians outside the camera’s view. This contrasts with NeRF’s ray marching process, which samples points along camera rays from near to far bounds and is generally more computationally intensive.

After projection, each 3D Gaussian becomes a 2D Gaussian (ellipse) on the image plane. A pixel-wise rendering algorithm \citep{zheng2024gps} then computes the color of each pixel by blending contributions from all overlapping Gaussians, sorted by their distance to the camera. The final pixel color $\bm{C}$ is computed via $\alpha$-blending: 
\begin{equation}
    \bm{C} = \sum_{i=1}^{|\mathcal{S}|} \bm{c}_i \alpha'_i \prod_{j=1}^{i-1}(1-\alpha'_j),
\end{equation}
where $\bm{c}_i$ is the color of Gaussian $i$, and $\alpha'_i$ is its pixel-wise opacity on the image plane, computed as: 
\begin{equation}
    \alpha'_i = \alpha_i \cdot e^{-\frac{1}{2}(\bm{x}'-\bm{\mu}'_i)^T \bm{\Sigma}_i^{'-1}(\bm{x}'-\bm{\mu}'_i)}.
\end{equation}
Here, $\bm{x}'$ is the pixel location, and $\bm{\mu}'_i$ and $\bm{\Sigma}'_i$ are the center and covariance of Gaussian $i$ in 2D image space.

To support real-time performance in high-resolution scenes, parallel computation techniques such as tile-based rasterization \citep{lassner2021pulsar} and CUDA-based rendering \citep{kerbl20233d} are utilized. Additionally, during training, density control techniques \citep{rota2024revising}, including point densification and point pruning, are applied to adjust the spatial distribution of Gaussians based on gradient and opacity information, further improving reconstruction quality.

A commonly used loss function combines the $\mathcal{L}_1$ color reconstruction loss and the Structural Dissimilarity Index Measure (D-SSIM) loss \citep{kerbl20233d}: 
\begin{equation}
    \mathcal{L} = (1-\lambda)\mathcal{L}_1 + \lambda \mathcal{L}_{D-SSIM},
\end{equation}
where $\mathcal{L}_1 = \frac{1}{N}\sum_{q\in \text{image}}||\bm{C}(q) - \hat{\bm{C}}(q)||_1$ is used to optimize the color accuracy, and $ \mathcal{L}_{D-SSIM} = \frac{1-SSIM}{2}$ is used to improve the structural similarity between the synthetic and true images.

The training pipeline for 3DGS is illustrated in Figure ~\ref{fig:traditional_method}(e) and involves the following steps: 
\begin{enumerate} 
    \item Capture multi-view images of the scene. Initialize the 3D Gaussian set using a point cloud generated from SfM or through random initialization. 
    \item Project the 3D Gaussians onto the image plane and generate synthetic views using the rendering process. 
    \item Optimize Gaussian parameters by minimizing the difference between synthetic and ground truth images, while dynamically refining the distribution of Gaussians based on learned gradients and opacity. 
\end{enumerate}

\subsection{Metrics for evaluating 3D reconstruction}  \label{sec:2.4}
Evaluating 3D reconstruction quality is crucial for reliable plant phenotyping, supporting tasks such as trait analysis, yield estimation, and growth monitoring. In this survey, we categorize evaluation metrics into three main groups: pixel-level image quality metrics, geometry-level structural metrics, and trait estimation metrics. Table~\ref{tab:evaluation_metrics} summarizes these metrics, highlighting their mechanisms, ideal outcomes, and relevance to plant phenotyping.

\begin{table*}[]
\centering
\renewcommand{\arraystretch}{1.6}
\caption{Summary of evaluation metrics in 3D reconstruction. In the table, the arrows indicate the preferred direction for each metric: ``$\uparrow$'' means higher values are better, ``$\downarrow$'' indicates that lower values are preferred, and ``$\rightarrow 1$'' denotes that values closer to 1 are ideal.}
\label{tab:evaluation_metrics}
\resizebox{0.95\textwidth}{!}{%
\begin{tabular}{|l|l|l|}
\hline
\textbf{Metrics}           & \textbf{Mechanisms}                                                            & \textbf{Key phenotypic focus}                              \\ \hline \hline
PSNR ($\uparrow$)             & Logarithmic ratio of peak signal power to MSE (in decibels)                         & Preservation of fine textures and surface details                 \\
SSIM ($\rightarrow 1$)             & Compares luminance, contrast, and structural similarity                        & Retention of overall structural consistency                      \\ 
LPIPS ($\downarrow$)           & Perceptual similarity based on deep feature embeddings                                                & Realistic rendering of texture and visual appearance          \\ \hline
IoU ($\uparrow$)              & Overlap ratio between predicted and ground truth segmentation masks                                     & Completeness of plant structure reconstruction                \\ 
BO ($\uparrow$)              & Overlap between predicted and true object boundaries                                          & Accuracy in preserving fine structural contours                     \\
Chamfer Distance ($\downarrow$) & Mean bidirectional nearest-neighbor distance between point clouds                & Geometric fidelity of plant structures                    \\ 
Precision ($\uparrow$)        & Ratio of true positives among all predicted positives                                 & Reduction of false positives in results   \\ 
Recall ($\uparrow$)           & Ratio of true positives among all actual positives                                     & Coverage of true plant structures                       \\ 
F1-Score ($\uparrow$)         & Harmonic mean of precision and recall                                                & Balanced performance in detecting plant features          \\ \hline
$R^2$ ($\rightarrow 1$)            & Proportion of variance explained in regression models                          & Accuracy of estimated traits (e.g., height, leaf area)      \\
RMSE ($\downarrow$)            & Root mean squared error between predictions and ground truth                   & Pixel-level reconstruction accuracy                       \\ 
MAPE ($\downarrow$)             & Mean absolute percentage error in trait predictions                                    & Accuracy of quantitative trait estimation                 \\ \hline
\end{tabular}%
}
\end{table*}

\subsubsection{Pixel-level metrics}
Pixel-level metrics assess 2D renderings of reconstructed 3D models by comparing them with ground truth images.

\textit{Peak Signal-to-Noise Ratio (PSNR):}  
Derived from the mean square error (MSE), PSNR provides a logarithmic measure of reconstruction quality, and higher PSNR values imply finer details:
\begin{equation}
\text{PSNR} = 10 \cdot \log_{10}\!\Bigl(\frac{\text{MAX}_I^2}{\text{MSE}}\Bigr),
\end{equation}
where \(\text{MAX}_I\) is the maximum possible pixel value. In 3D plant reconstruction, PSNR is commonly used to assess the fidelity of reconstructed surfaces by measuring how closely the generated textures and geometric details resemble the ground truth \citep{zhao2024exploring}. 

\textit{Structural Similarity Index Measure (SSIM):}
SSIM evaluates the perceptual quality of reconstructed images by comparing luminance, contrast, and structural similarity with reference images:
\begin{equation}
\text{SSIM}(x,y) = \frac{(2\mu_x\mu_y + C_1)(2\sigma_{xy} + C_2)}{(\mu_x^2+\mu_y^2+C_1)(\sigma_x^2+\sigma_y^2+C_2)},
\end{equation}
where \(\mu_x, \mu_y\) denote the means, \(\sigma_x^2, \sigma_y^2\) are the variances, and \(\sigma_{xy}\) the covariance of images \(x\) and \(y\). Higher SSIM values indicate improved preservation of visual consistency in reconstructed surfaces \citep{zhao2024exploring}. 

\textit{Learned Perceptual Image Patch Similarity (LPIPS):}  
LPIPS leverages deep neural network features to compare perceptual similarity:
\begin{equation}
\text{LPIPS}(x,y)=\sum_{l}\frac{1}{H_lW_l}\sum_{h,w}\|w_l \odot(\hat{x}_{hw}^{l}-\hat{y}_{hw}^{l})\|_2^2,
\end{equation}
where \(\hat{x}_{hw}^{l}\) and \(\hat{y}_{hw}^{l}\) are the features from layer \(l\) of a deep network, and \(w_l\) are learned weights. Lower LPIPS scores imply stronger visual resemblance \citep{chopra2024agrinerf}.

\subsubsection{Geometry-level metrics}
Geometry-level metrics evaluate the structural fidelity of 3D models by comparing reconstructed geometry to reference models. They are essential for ensuring morphological accuracy and structural completeness.

\textit{Intersection over Union (IoU):}  
IoU measures the overlap between the predicted segmentation mask \(P\) and the ground truth \(G\), evaluating how well the reconstructed structure aligns with its reference:
\begin{equation}
\text{IoU} = \frac{|P \cap G|}{|P \cup G|}.
\end{equation}
Higher IoU indicates more accurate shape reconstruction \citep{shen2024biomass}.

\textit{Chamfer Distance (CD):}  
CD quantifies the discrepancy between two point clouds by computing the average bidirectional nearest-neighbor distance between the reconstructed (\(S_1\)) and reference (\(S_2\)) point clouds:
\begin{equation}
CD(S_1, S_2) = \frac{1}{|S_1|} \sum_{x \in S_1} \min_{y \in S_2} \|x - y\|_2^2 + \frac{1}{|S_2|} \sum_{y \in S_2} \min_{x \in S_1} \|x - y\|_2^2.
\end{equation}
Lower CD values indicate greater alignment, signifying improved reconstruction accuracy \citep{saeed2023peanutnerf}.

\textit{Boundary Overlap (BO):}  
BO assesses the correspondence between the boundaries extracted from the reconstructed model and those in the ground truth segmentation. Higher BO values indicate precise contour alignment, which is critical in delineating fine-grained plant structures. BO is mathematically defined as:  
\begin{equation}
BO = \frac{|E_p \cap E_g|}{|E_p \cup E_g|},
\end{equation}
where \(E_p\) represents the set of predicted edge pixels, \(E_g\) represents the set of ground truth edge pixels, ``\(\cap\)'' denotes the intersection, and ``\(\cup\)'' denotes the union \citep{yang2024paniclenerf}. 

\textit{Precision, Recall, and F1-Score:}  
These metrics evaluate segmentation and detection tasks in 3D plant models, measuring the balance between accuracy and completeness:
\begin{align}
\text{Precision} &= \frac{\text{TP}}{\text{TP}+\text{FP}}, \quad
\text{Recall} = \frac{\text{TP}}{\text{TP}+\text{FN}}, \\
\text{F1-Score} &= 2 \cdot \frac{\text{Precision}\times \text{Recall}}{\text{Precision}+\text{Recall}},
\end{align}
where TP, FP, and FN represent true positives, false positives, and false negatives, respectively \citep{shen2024biomass, yang2024paniclenerf}. 

\subsubsection{Trait-specific metrics}
To ensure the reconstructed models are useful for phenotypic analysis, trait-level metrics are used to evaluate prediction accuracy of plant features (e.g., height, biomass, leaf area).

\textit{Coefficient of Determination (\(R^2\)):}  
This metric quantifies the proportion of variance in plant traits (e.g., height, leaf area) explained by the reconstruction:
\begin{equation}
R^2 = 1 - \frac{\sum_{i=1}^{N}(y_i - \hat{y}_i)^2}{\sum_{i=1}^{N}(y_i - \bar{y})^2}.
\end{equation}
Values closer to 1 suggest that the model accurately predicts phenotypic traits, validating the reconstruction’s utility for quantitative analysis \citep{zhu2024three, yang2024paniclenerf}.

\textit{Root Mean Squared Error (RMSE):}  
RMSE quantifies the average squared difference between predicted and actual values, assessing the accuracy of relative estimation:
\begin{equation}
\text{RMSE} =\sqrt{\frac{1}{N}\sum_{i=1}^{N}\bigl({y_i - \hat{y}_i}\bigr)^2.}
\end{equation}
Lower RMSE values indicate better predictive performance, making it a crucial metric in biomass estimation, where accurate trait predictions are essential for phenotypic analysis \citep{yang2024paniclenerf}.

\textit{Mean Absolute Percentage Error (MAPE):}  
MAPE measures the average absolute percentage error between the predicted and true trait values:
\begin{equation}
\text{MAPE} = \frac{1}{N}\sum_{i=1}^{N}\left|\frac{y_i - \hat{y}_i}{y_i}\right| \times 100\%.
\end{equation}
Lower MAPE values indicate that the reconstruction produces robust trait predictions \citep{choi2024nerf}.

\section{Current Progress of 3D Reconstruction in Plant Phenotyping}
\label{sec3}
In this section, we will review the latest advancements in 3D reconstruction techniques applied to plant phenotyping, highlighting key research contributions that have shaped the field. By delving into both theoretical innovations and practical implementations, we aim to provide a comprehensive overview of how these methods enhance the understanding of plant morphology and enable automated, high-precision phenotyping.

\begin{table*}[]
\centering
\renewcommand{\arraystretch}{1.6}
\caption{Summary of recent representative reconstruction research in plant phenotyping.}
\label{tab:apps}
\resizebox{0.98\textwidth}{!}{%
\begin{tabular}{|l|l|l|l|l|l|}
\hline
\textbf{Model Type}                & \textbf{Reference}    & \textbf{Crop}                                             & \textbf{Method}                & \textbf{Downstream Tasks}            & \textbf{Devices}                        \\ \hline \hline
\multirow{3}{*}{Classical Methods} & \cite{gene2020fruit}     & Apple                                                  & VisualSFM                      & Fruit detection and yield prediction       & LiDAR                                  \\
                    & \cite{zhang2024cucumber}     & Cucumber                                                  & VisualSFM                      & Cucumber seedling segmentation       & RGB-D                                  \\
                    & \cite{wang20223dphenomvs}    & Tomato                                                    & 3DPhenoMVS                     & Phenotypic trait analysis            & Digital camera                          \\ 
                                   & \cite{chen2023point}         & Chinese Cabbage & PF-Net                         & Leaf reconstruction                  & Azure Kinect camera                     \\ \hline
\multirow{8}{*}{NeRF}              & \cite{zhao2024exploring}     & Pepper                                                    & Instant-NSR                    & 3D fruit segmentation                & GoPro Hero 11                           \\ 
                                   & \cite{chopra2024agrinerf}    & Apple, peach                                              & EvDeblurNeRF                              & Fruit detection                      & -                                      \\ 
                                   & \cite{hu2024high}            & Pitahaya, grape, orange, etc.                            & Instant-NGP, Instant-NSR       & Phenotypic reconstruction            & -                                       \\  
                                   & \cite{yang2024paniclenerf}   & Rice                                                      & Instant-NGP                    & Rice panicle segmentation  & Smartphone                              \\ 
                                   & \cite{zhu2024three}          & Potted plants                                             & SFM-NeRF                       & Plant organ analysis                 & Professional camera                     \\  
                                   & \cite{choi2024nerf}          & Tomato                                                    & Nerfacto                       & Measurements of crop parameters      & RGB camera                              \\  
                                   & \cite{arshad2024evaluating}  & Maize                                                     & Instant-NGP, TensoRF, Nerfacto & Plant Geometry Reconstruction        & iPhone 13 Pro \\ 
                                   & \cite{saeed2023peanutnerf}   & Peanut                                                    & Nerfacto                       & Peanut pod detection                 & -                                       \\ 
                                    & \cite{zhang2024neural}   & Strawberry                                                    & Nerf                       & Strawberry 3D scene reconstruction and rendering                 & Huawei smartphone                                \\ \hline
\multirow{3}{*}{3DGS}              & \cite{jiang2024estimation}   & Cotton                                                    & 3DGS                           & Cotton Boll and stem length analysis & iPhone 11                               \\ 
                                   & \cite{ojo2024splanting}      & Canola, wheat, bean, etc.                & 3DGS                           & Plant shoot architecture             & -                                       \\ 
                                   & \cite{shen2025plantgaussian} & Wheat, tobacco, corn                                      & 3DGS                           & 3D plant visualization               & Smartphone                             \\ \hline
\end{tabular}%
}
\end{table*}

\subsection{Classical methods}
Classical methods, including both active and passive approaches, have been extensively studied in the literature, as summarized in Table~\ref{tab:reviews}. Therefore, this subsection highlights only a few representative methods. For a more comprehensive review of these classical 3D reconstruction methods, readers are referred to the studies listed in Table~\ref{tab:reviews}.

Active reconstruction methods using depth sensors (see Section~\ref{sec:2.1.1}) are among the most commonly used approaches in plant phenotyping. For instance, in \cite{gene2020fruit}, a LiDAR-based system was developed to reduce fruit occlusions—common in orchard environments that lead to yield underestimation—by combining forced air flow (via an air-assisted sprayer) and multi-view sensing. The system was evaluated in a commercial Fuji apple orchard, detecting over 80\% of visible fruits and predicting yield with an RMSE below 6\%. These strategies also improved fruit detection by 6.7\% and 6.5\%, respectively, and enabled accurate geometric characterization of canopy traits such as height, width, cross-sectional area, and leaf area.
Additionally, in \cite{chen2023point}, a Point Fractal Network-based method was proposed to complete incomplete point clouds of flowering Chinese cabbage caused by occlusions and the limitations of missing and incomplete data caused by 3-dimensional point cloud technology. The study constructed a dedicated point cloud dataset of cabbage leaves and used deep learning to recover missing structures under various occlusion scenarios. A novel single-view RGB-D framework was also introduced to complete localized gaps in leaf point clouds captured by RGB-D cameras. Experimental results showed that prior to point cloud completion, leaf area estimation yielded an $R^2$ of 0.9162, RMSE of 15.88 cm², and average relative error of 22.11\%. After completion, performance improved significantly, achieving an $R^2$ of 0.9637, RMSE of 6.79 cm², and average relative error of 8.82\%. This approach notably enhances the accuracy of non-invasive phenotypic parameter estimation, offering a robust solution for 3D plant analysis.

Passive reconstruction methods, such as SfM (see Section~\ref{sec:2.1.2}), have also been widely explored in plant phenotyping. In \cite{wang20223dphenomvs}, a 3D phenotyping pipeline named 3DPhenoMVS was proposed to estimate 17 phenotypic traits of tomato plants across their entire life cycle. The pipeline utilized 3D structural data generated from multi-view images using SfM algorithms \citep{schonberger2016structure}. Among the 17 traits, six were selected for accuracy evaluation based on the availability of corresponding ground truth measurements obtained manually. The results, implemented with the free academic software VisualSFM \citep{wu2013towards}, showed strong agreement, with $R^2$ values ranging from 0.72 to 0.97 between the estimated and manually measured traits. Additionally, to study the environmental impact on tomato growth and yield in greenhouse conditions, eight tomato plants were monitored across seven growth stages under varying light intensities, temperatures, and humidity levels. The findings indicated that stronger light intensity, coupled with moderate temperature and humidity, led to higher biomass accumulation and improved yield.
Similarly, VisualSFM \citep{weidner2021influence}, an enhanced SfM method and open-source tool for reconstructing 3D structures by analyzing geometric relationships between images, was employed to generate 3D models of cucumber seedlings for downstream semantic segmentation tasks. Extensive experiments on a cucumber seedling dataset demonstrated that the proposed SN-MGGE network outperformed several mainstream segmentation models (e.g., PointNet++, PointMLP, and AGConv), achieving a mean mIoU of 94.90\% and an overall accuracy (OA) of 97.43\%.

\subsection{NeRF in plant phenotyping}

\begin{figure*}[!ht]
  \centering
  \includegraphics[width=0.9\textwidth]{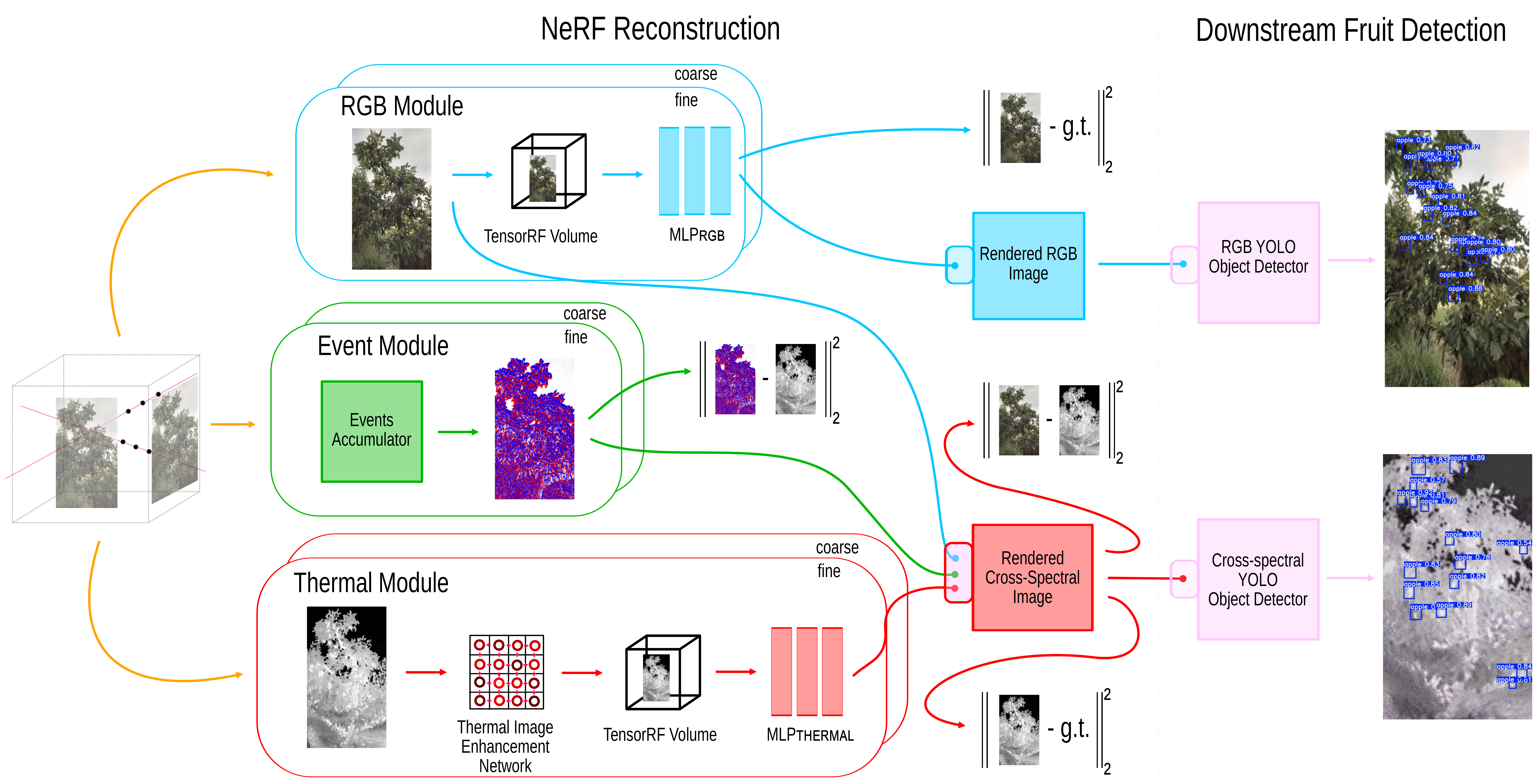}
  \vspace{1pt}
  \caption{Architecture of AgriNeRF and downstream fruit detection. Adapted from \cite{chopra2024agrinerf}.
  }
  \label{fig:AgriNeRF}
\end{figure*}

NeRF techniques, as elaborated in Section 2.2, have fundamentally transformed the design, development, and deployment phases of 3D reconstruction technologies. These cutting-edge methods are increasingly being employed by researchers to advance various plant phenotyping applications, such as high-throughput trait analysis \citep{yang2024paniclenerf}, structural modeling of plant architecture \citep{arshad2024evaluating}, and precision agriculture optimization \citep{zhao2024exploring}. By facilitating highly accurate 3D reconstructions and photorealistic visualizations, NeRF empowers more robust and automated phenotypic assessments, significantly enhancing the efficiency, scalability, and precision of plant research efforts.  

Notably, researchers have extended NeRF’s applications to orchard environments, demonstrating its adaptability to complex agricultural settings. For example, in \cite{hu2024high}, two state-of-the-art NeRF methods were explored for the 3D reconstruction of crop and plant models: Instant-NGP \citep{muller2022instant}, known for its ability to generate high-quality images with remarkable training and inference speed, and Instant-NSR \citep{zhao2022human}, which enhances reconstructed geometry by incorporating the Signed Distance Function (SDF) \citep{wang2021neus} during training. To test their framework, the authors collected a novel plant phenotype dataset comprising real plant images from production environments. Experimental results showed that NeRF achieved impressive performance in synthesizing novel-view images and produced reconstruction results competitive with RealityCapture\footnote{\url{https://www.capturingreality.com/}}, a leading commercial software for 3D Multi-View Stereo (MVS)-based reconstruction.
Additionally, in \cite{chopra2024agrinerf}, the authors proposed an innovative multi-modal NeRF framework that reconstructs orchard scenes using data from three distinct camera types: RGB, thermal, and event cameras (Figure~\ref{fig:AgriNeRF}). To achieve this, they introduced a regularization loss term designed to maximize feature correlation across RGB, event, and thermal frames. This approach enabled the training of a cross-spectral reconstruction of the scene, effectively integrating features from both the visible and infrared bands of the electromagnetic spectrum. To enhance reconstruction quality, the framework incorporated EvDeblurNeRF \citep{cannici2024mitigating} for scene reconstruction and employed YOLOv8 \citep{reis2023real} for downstream fruit detection tasks. The framework was rigorously evaluated on real-world multi-modal datasets collected from diverse environments, including apple and peach orchards, as well as gardens and parks. The results demonstrated a remarkable +44.8\% increase in mAP50 for fruit detection, underscoring the method’s superior performance in capturing detailed and accurate 3D representations. 

\begin{figure*}[!ht]
  \centering
  \includegraphics[width=0.8\textwidth]{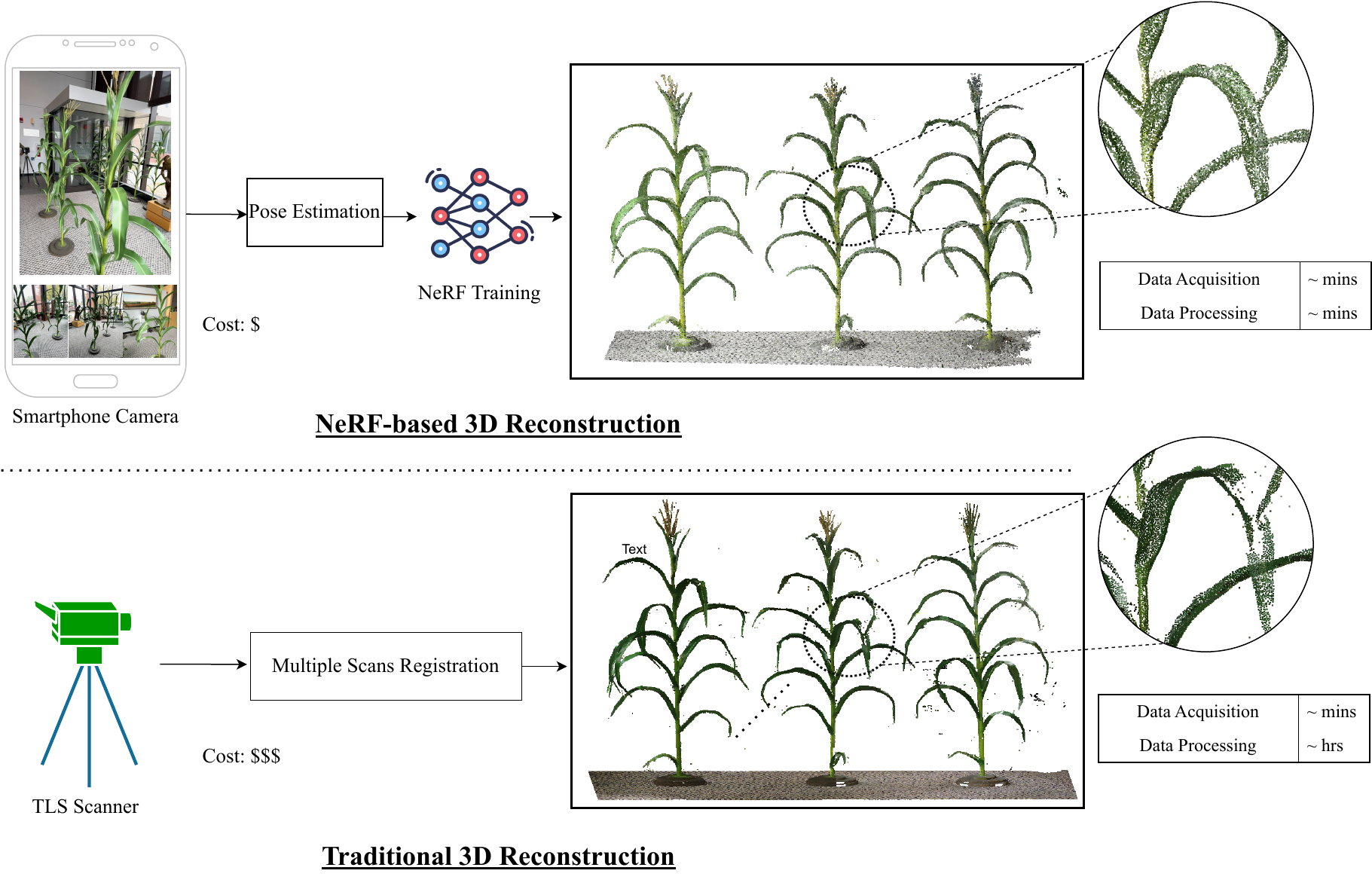}
  \vspace{1pt}
  \caption{Workflow comparison of NeRF-based and traditional 3D reconstruction methods for corn plants. Adapted from \cite{arshad2024evaluating}.
  }
  \label{fig:nerf_corn}
\end{figure*}

Beyond orchards, NeRF has also been applied to other crop species, such as rice \citep{yang2024paniclenerf}, maize \citep{arshad2024evaluating}, peanut \citep{saeed2023peanutnerf}, and strawberry \citep{zhang2024neural}, further demonstrating its versatility and potential in diverse agricultural contexts. For example, in \cite{yang2024paniclenerf}, the authors proposed a novel framework, PanicleNeRF, built on Instant-NGP \citep{muller2022instant}, to enable high-precision and low-cost 3D reconstruction of rice panicles in field conditions using video captured by a smartphone. To support accurate reconstruction, the method integrated the large-scale Segment Anything Model (SAM) \citep{kirillov2023segment} with YOLOv8 \citep{reis2023real} for precise segmentation of rice panicle images, thereby enhancing the quality of the extracted features and improving the overall reconstruction performance. Experimental results demonstrated that PanicleNeRF effectively addressed the 2D image segmentation task, achieving a mean F1 score of 86.9\% and a mean Intersection over Union (IoU) of 79.8\%, highlighting its strong performance in complex field environments. 
In \cite{arshad2024evaluating}, three corn scenarios with increasing levels of complexity, single corn plant indoors, multiple corn plants indoors, and multiple corn plants in a field with other plants, were reconstructed and evaluated using Nerfacto \citep{tancik2023nerfstudio} to assess reconstruction fidelity. The quality of the NeRF-based reconstructions was further validated by comparing them against point clouds obtained from LiDAR scans, which served as the ground truth. In the most realistic field scenario, the NeRF model achieved an F1 score of 74.6\% after 30 minutes of GPU training, demonstrating the effectiveness of NeRF-based methods for 3D reconstruction in complex and challenging agricultural environments. The proposed framework is shown in Figure~\ref{fig:nerf_corn}.
Additionally, in \cite{saeed2023peanutnerf}, Nerfacto \citep{tancik2023nerfstudio} was utilized for the 3D reconstruction of peanut plants from captured 2D images. To enable 3D detection of peanut pods, the study employed Frustum PVCNN \citep{liu2019point}, leveraging frontal-view reconstructions along with corresponding 2D images. The pod detection achieved a precision of approximately 0.7 at an IoU threshold of 0.5 on the validation set, demonstrating the potential of NeRF-based methods for accurate pod localization in peanut crops.

\begin{figure*}[!ht]
  \centering
  \includegraphics[width=0.8\textwidth]{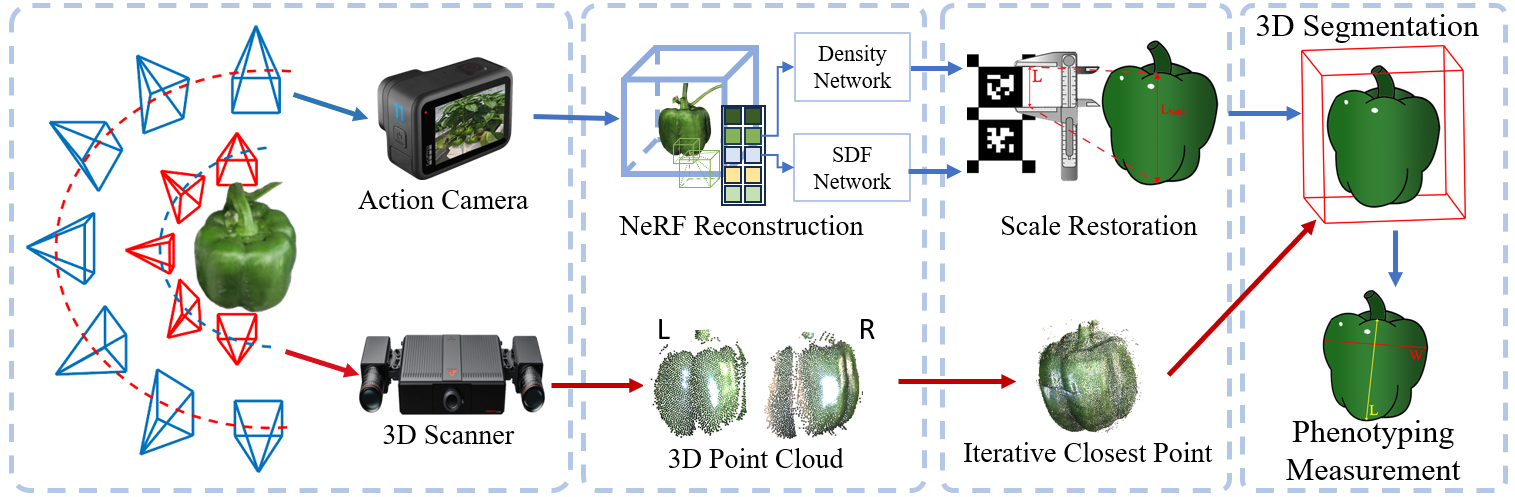}
  \vspace{1pt}
  \caption{Framework for 3D phenotyping for bell pepper using NeRF-based reconstruction and 3D scanning \cite{zhao2024exploring}. The process integrates action camera and 3D scanner data, followed by NeRF reconstruction, scale restoration, segmentation, and phenotypic measurements. 
  }
  \label{fig:nerf_pepper}
\end{figure*}

NeRF has also been applied in greenhouse and laboratory environments, further extending its utility in controlled settings for plant phenotyping \citep{zhao2024exploring, zhu2024three, choi2024nerf}. These environments offer unique advantages, such as consistent lighting conditions and minimized occlusions, which allow for more precise 3D reconstructions of plant structures \citep{akhtar2024unlocking}. 
For instance, in \citep{zhao2024exploring}, Instant-NSR \citep{zhao2022human} was employed to achieve accurate 3D phenotyping of pepper plants in greenhouse environments. Figure~\ref{fig:nerf_pepper} shows the proposed framework. To evaluate its performance, traditional point cloud registration on 3D scanning data was implemented for comparison. Experimental results demonstrated that NeRF achieved competitive accuracy compared to conventional 3D scanning methods, with a mean distance error of 0.865 mm between the scanner-based and NeRF-based reconstructions, highlighting its potential as a reliable tool for precise plant phenotyping. 
Additionally, in \cite{zhu2024three}, a 3D phenotyping pipeline was proposed to acquire detailed, non-destructive information from living potted plants by integrating NeRF with path analysis \citep{vicari2019leaf}. An indoor collection system was developed to capture multi-view image sequences of potted plants, enabling precise 3D reconstruction. To achieve this, SfM \citep{dellaert2000structure} and NeRF \citep{mildenhall2021nerf} were combined to generate 3D point clouds, which were subsequently denoised and calibrated for enhanced accuracy.  The system extracted key phenotypic parameters of plant organs, including height, stem thickness, leaf length, leaf width, and leaf area, which were validated through manual measurements. Experimental results demonstrated strong correlations between the extracted traits and ground truth values, with coefficient of determination ($R^2$) values ranging from 0.89 to 0.98, underscoring the accuracy and reliability of the proposed pipeline for high-quality 3D plant reconstruction and phenotypic assessment.
Moreover, in \cite{choi2024nerf}, an automated RGB image acquisition framework was developed using unmanned greenhouse robots integrated with Nerfacto \citep{tancik2023nerfstudio} to generate dense point clouds for non-destructive and precise measurement of crop parameters. This system was designed to assess tomato crops in greenhouse conditions, enabling accurate estimation of key traits such as node length, leaf area, and fruit volume with minimal manual intervention. The results showed $R^2= 0.973$ and MAPE = 0.089 for inter-node length measurements, while segmented leaf point clouds and reconstructed meshes achieved $R^2 = 0.953$ and MAPE = 0.090 for leaf area measurements.

\begin{figure}[!ht]
  \centering
  \includegraphics[width=0.34\textwidth]{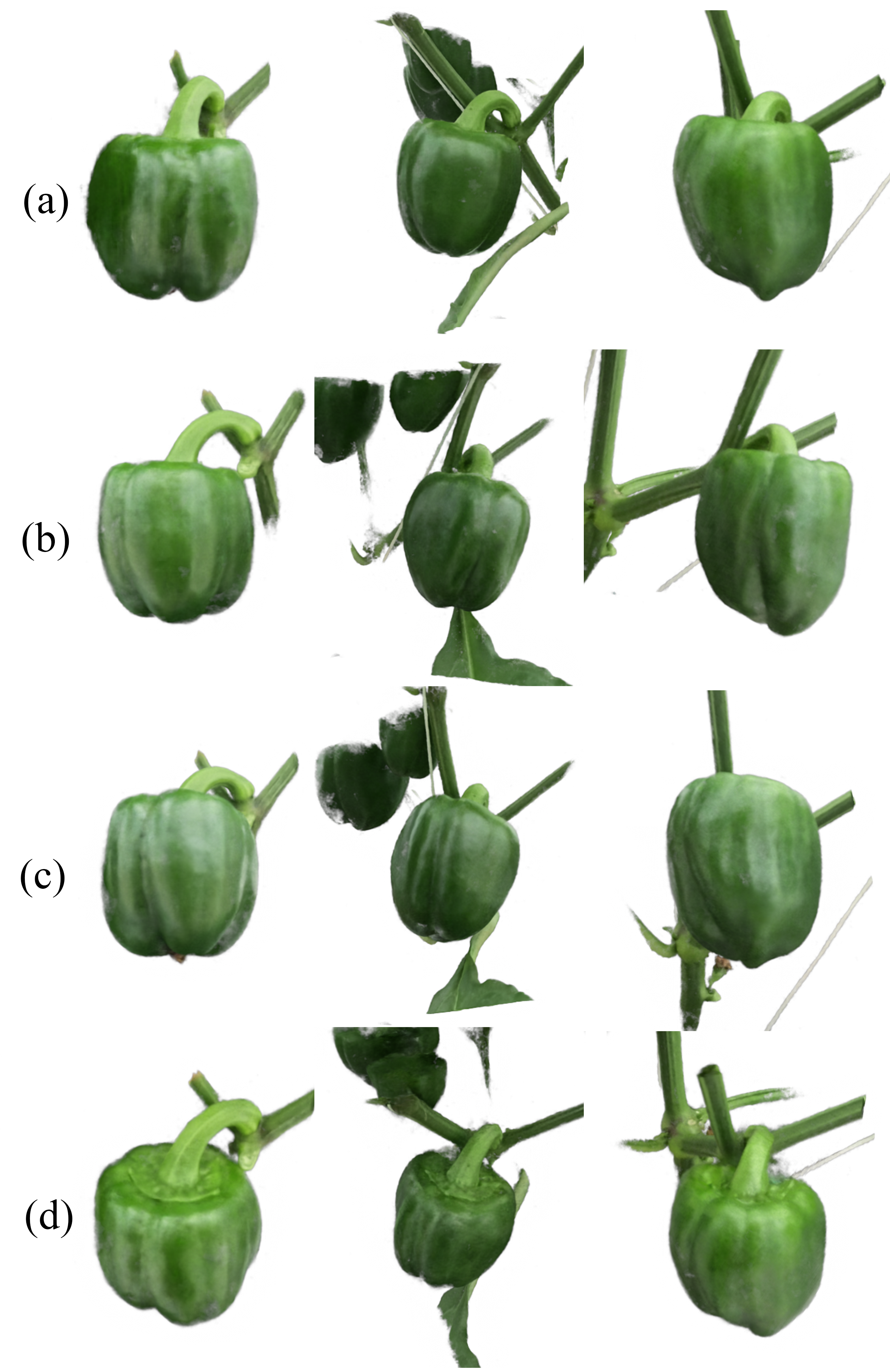}
  \vspace{-1pt}
  \caption{Multi-view renderings of the NeRF model (a) Frontviews (b) Side views (c) Top views (d) Elevation views \citep{zhao2024exploring}. }
  \label{fig:pepper_rendering}
\end{figure}

\subsection{3DGS in plant phenotyping}
While NeRF-based methods have demonstrated success in 3D plant reconstruction, they often suffer from high computational costs, long training times, and slow rendering speeds, making them less suitable for real-time applications in precision agriculture \citep{bao20253d, chen2024survey}. 3DGS has emerged as a promising alternative, offering faster training, real-time rendering, and high reconstruction fidelity while maintaining fine-grained plant structural details \citep{kerbl20233d}. Due to these advantages, 3DGS has garnered increasing attention from the agricultural research community for its potential in high-throughput phenotyping and scalable crop monitoring. 

\begin{figure}[!ht]
  \centering
  \includegraphics[width=0.45\textwidth]{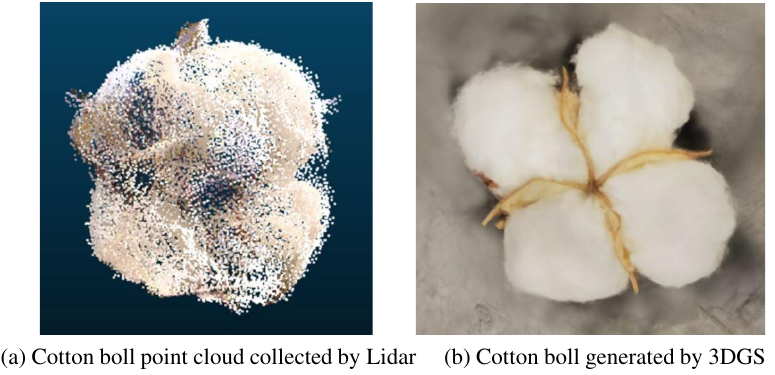}
  \vspace{-1pt}
  \caption{Examples of 3D cotton bolls generated by Lidar and 3DGS \citep{jiang2024estimation}. }
  \label{fig:3dgs_cotton}
\end{figure}

Notably, in \cite{jiang2024estimation}, a 3DGS-based approach was proposed for cotton phenotypic trait measurement using instance segmentation. The 3DGS model for the cotton plant scene was trained on sparse point clouds generated from 2D images processed with COLMAP \citep{schonberger2016structure}. To segment objects of interest, the YOLOv8x instance segmentation model \citep{reis2023real} was employed to produce 2D masks. These masks were then integrated into Segment Any 3D Gaussians (SAGA) \citep{cen2023segment}, enabling the extraction of individual 3DGS models for cotton bolls and the main stem. Finally, key cotton phenotypic traits, including boll count and main stem length, were estimated. Comparative analysis with ground truth measurements showed that the method achieved a mean absolute percentage error (MAPE) of 11.43\% for cotton boll counting and 10.45\% for main stem length, demonstrating the effectiveness of 3DGS in precise, non-destructive plant phenotyping. Figure~\ref{fig:3dgs_cotton} shows the generated 3D cotton bolls by the 3DGS model and Lidar, respectively.

In \cite{ojo2024splanting}, 3DGS was evaluated for capturing high-fidelity 3D plant representations. To facilitate efficient data acquisition and processing, the authors developed Splants, a specialized 3DGS-based framework designed for fast and simple foreground object capture. Additionally, the Splants dataset was introduced, including a diverse range of agronomically important plant species, growth stages, and environmental backgrounds, with the expectation of continued expansion to support broader applications in plant phenotyping and precision agriculture. Specifically, the dataset includes nine plant species, such as canola, wheat, bean, chickpea, lentil, and pea, providing a valuable resource for high-throughput 3D plant analysis and trait extraction. To validate that the optimized splats accurately represent plant structures, the study explored various analytical approaches. These included skeletonizing Splants using splat locations to reconstruct plant architectures, computing plant volume, isolating canola flowers through splat radiance, and differentiating plant organs by analyzing splant orientation within splat clusters. These findings highlight the potential of 3DGS-based plant modeling for detailed structural analysis and phenotypic trait measurement in agricultural research.

\begin{figure*}[!ht]
  \centering
  \includegraphics[width=0.8\textwidth]{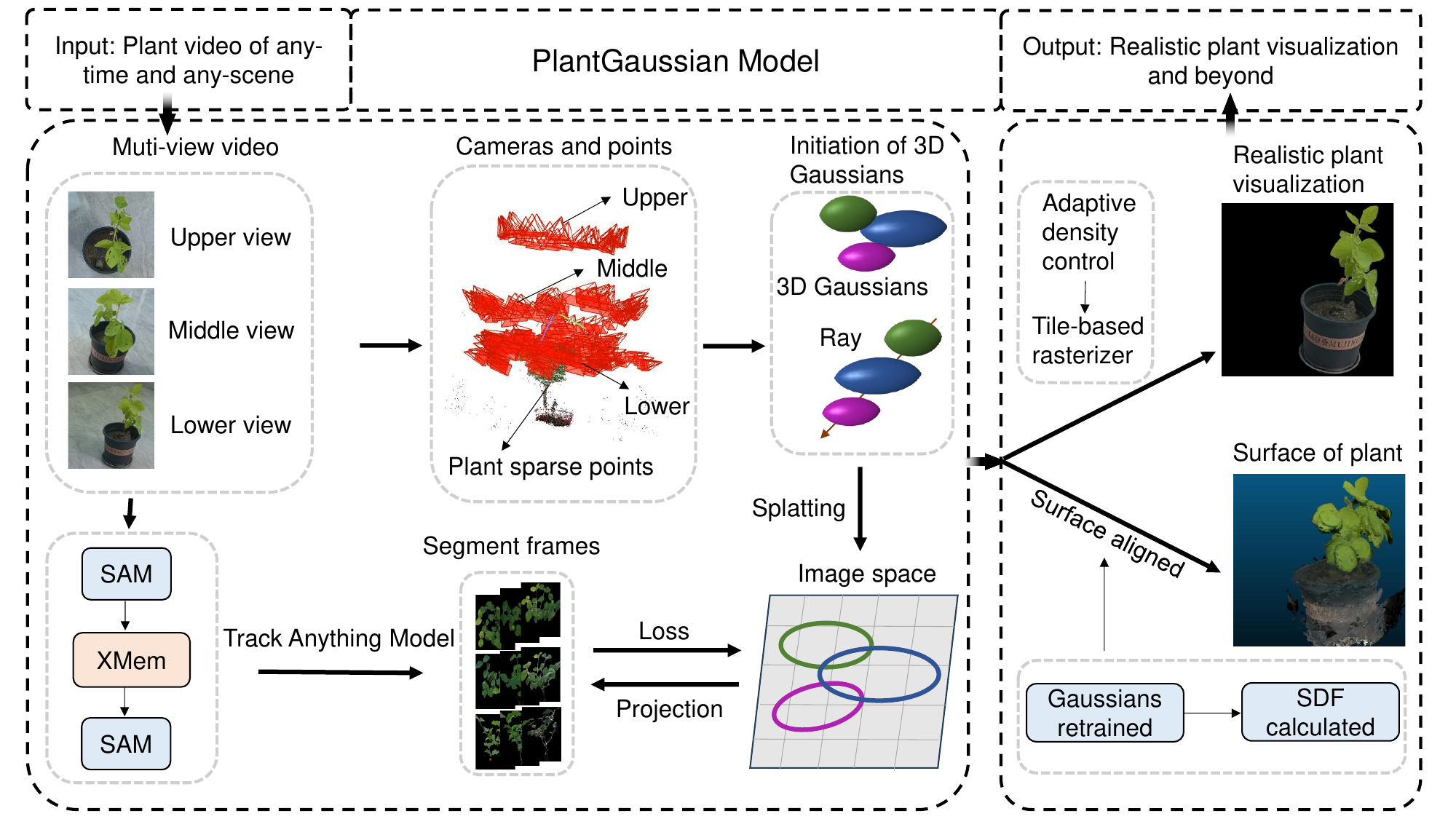}
  \vspace{1pt}
  \caption{The framework of the PlantGaussian model \citep{shen2025plantgaussian}. The model processes multi-view videos to generate plant visualizations and surface outputs, starting with sparse reconstruction to derive sparse points and camera poses. Next, an improved Track Anything Model \citep{yang2023track} isolates the plant foreground, which, combined with the sparse points and poses, initializes 3D Gaussian points.}
  \label{fig:3dgs_plant}
\end{figure*}

In \cite{shen2025plantgaussian}, the authors introduced PlantGaussian, a framework for generating realistic 3D visualizations of plants across different time points and environmental conditions. To address the limitations of traditional Gaussian reconstruction techniques in complex planting environments, the framework integrated the Segment Anything Model (SAM) \citep{kirillov2023segment} and tracking algorithms to enhance plant segmentation and temporal consistency. A novel mesh partitioning technique was employed to convert Gaussian-rendered outputs into measurable plant meshes, providing a robust methodology for accurate 3D plant morphology phenotyping. Additionally, the PlantGaussian dataset was developed, featuring images of four crop species: corn, wheat, soybean, and tobacco, captured under diverse conditions and growth stages. The framework enabled high-fidelity plant visualization and 3D mesh reconstruction for plant morphological phenotyping, requiring only plant image sequences as input, making it a scalable and efficient solution for 3D plant structural analysis. The framework is shown in Figure~\ref{fig:3dgs_plant}.

\section{Discussions}
\label{sec4}
This section outlines the limitations of current research in NeRF and 3DGS-based plant phenotyping and explores potential solutions. We also discuss future directions, including multi-modal reconstruction, hyperspectral imaging, editable models, and advanced computer vision integration, to enhance scalability, accuracy, and efficiency in plant phenotyping.

\subsection{Limitations of current research}
Despite significant advancements in NeRF and 3DGS-based plant phenotyping, several challenges remain which limit their practical implementation in real-world agricultural settings. First, NeRF models are computationally expensive, requiring long training times and significant GPU resources, making them less suitable for real-time applications \citep{gao2022nerf}. While 3DGS offers faster rendering, its reconstruction accuracy may suffer in occluded or highly complex plant structures, leading to potential errors in morphological trait extraction \citep{chen2024survey, qin2024langsplat}. 
Current NeRF and 3DGS-based plant phenotyping applications are primarily built on classical models, limiting their adaptability to dynamic agricultural environments. However, recent advancements—such as NeRF models \citep{garbin2021fastnerf, lin2024fastsr} and 3DGS models \citep{chen2024mvsplat, cheng2024gaussianpro}, which significantly enhance reconstruction accuracy and efficiency—require further exploration and validation for broader adoption in high-throughput crop monitoring and precision agriculture.

Another limitation is the scarcity of large, well-annotated datasets for training and validating NeRF and 3DGS models. Current datasets, such as PlantGaussian \citep{shen2025plantgaussian} and Splants \citep{ojo2024splanting} for plant phenotyping, provide valuable benchmarks but are still limited in scale, species diversity, and environmental conditions. The lack of standardized evaluation metrics also makes it challenging to compare different NeRF and 3DGS-based approaches effectively. Recent advances in the general computer science domain highlight promising solutions to this challenge. For instance, ShapeSplat \citep{ma2024shapesplat} is a large-scale dataset of Gaussian splats, comprising 65K objects from 87 unique categories, designed for self-supervised pretraining. Such large-scale datasets have the potential to enhance model robustness and generalizability. Developing similar large-scale, diverse datasets tailored to plant phenotyping, combined with self-supervised learning techniques, could significantly improve 3D plant reconstruction accuracy, scalability, and cross-domain adaptability.

Furthermore, multi-view image acquisition remains a bottleneck for large-scale agricultural applications \citep{akhtar2024unlocking}. While robotic platforms and automated imaging systems have been explored \citep{zhao2024exploring}, issues such as camera calibration, motion blur, and inconsistent lighting introduce artifacts in 3D reconstructions. Future research should focus on adaptive data acquisition strategies, such as integrating Unmanned Aerial Vehicles (UAVs) for improved scalability and efficiency \citep{Turki_2022_CVPR}.

\subsection{Future research directions}
3D reconstruction with NeRF and 3DGS is a rapidly evolving field with significant potential for high-fidelity plant phenotyping. In this subsection, we highlight several key future directions, though many more opportunities remain for exploration and innovation.

\subsubsection{Multi-modal reconstruction}
Integrating multi-modal data sources can significantly enhance 3D plant reconstruction accuracy by leveraging complementary information from RGB, thermal, LiDAR, hyperspectral, and event cameras. For instance, incorporating depth sensors into NeRF and 3DGS pipelines can help mitigate errors in plant structure reconstruction \citep{zhang2024nerf}.
In \cite{chopra2024agrinerf}, the authors explored a multi-modal NeRF framework for reconstructing orchard scenes, utilizing RGB, thermal, and event cameras. This approach demonstrated promising results in orchard environments, highlighting the potential of multi-modal fusion for improved scene understanding and cross-spectral consistency. However, further research is required to optimize cross-spectral alignment, enhance sensor fusion techniques, and improve model robustness in dynamic agricultural settings.
Additionally, in \cite{lu2023large}, the authors introduced the OMMO dataset, a large-scale outdoor multi-modal dataset containing calibrated images, point clouds, and prompt annotations for complex objects and scenes. This dataset establishes a new benchmark for various NeRF-based outdoor tasks, including novel view synthesis, diverse 3D representation, and multi-modal NeRF applications. Developing similar work tailored to agricultural environments could significantly enhance the effectiveness of NeRF and 3DGS models in plant phenotyping and precision farming.

\subsubsection{3D hyperspectral reconstructions}
Hyperspectral imaging (HSI) has emerged as a powerful tool for non-invasive plant trait measurement, offering rich spectral information beyond traditional RGB images \citep{saric2022applications}. Integrating hyperspectral data with NeRF and 3DGS presents exciting opportunities for functional phenotyping, enabling the estimation of chlorophyll content, nutrient levels, and stress responses in plants. Existing 3D hyperspectral reconstruction methods typically rely on a combination of expensive hardware (e.g., hyperspectral cameras and scanning systems) and computationally intensive techniques for spectral data processing and interpretation \citep{behmann2016generation, saric2022applications}. However, these methods face limitations in spectral resolution preservation, reconstruction efficiency, expensive costs, and geometric accuracy when applied to complex plant structures.

In \cite{chen2024hyperspectral, zhang2024compressing, li2024spec}, researchers explored novel approaches to hyperspectral 3D reconstruction based on NeRF, incorporating deep learning-based spectral compression, multi-modal fusion, and optimized calibration techniques to enhance accuracy and computational efficiency in HSI-driven modeling. Similarly, \cite{thirgood2024hypergs, sinha2024spectralgaussians} introduced cross-spectral rendering framework based on 3DGS, enabling the generation of realistic and semantically meaningful Gaussian splats from registered multi-view spectral maps.
Future research should focus on self-supervised learning techniques to improve feature extraction from hyperspectral NeRF/3DGS models, as well as enhancing segmentation accuracy in complex agricultural phenotyping scenarios. Advancing cross-modal fusion techniques and real-time spectral rendering pipelines will be crucial for scalable, high-throughput 3D plant phenotyping.

\subsubsection{VR/AR extension}
VR/AR technologies allow researchers, agronomists, and educators to interact with highly detailed 3D plant models, providing a new dimension to data analysis, training, and decision-making \citep{cheng2024development}. The integration of NeRF and 3DGS-based 3D reconstructions with Virtual Reality (VR) and Augmented Reality (AR) presents exciting possibilities for immersive plant phenotyping and agricultural visualization \citep{yu2010research}. Notably, NeRF and 3DGS excel in photo-realistic quality and seamless, continuous view changes of 3D plant structures, enabling high-fidelity visualization without losing view-dependent effects \citep{deng2022fov, li2022rt}. This capability makes NeRF/3DGS-powered VR/AR environments particularly well-suited for interactive crop modeling, plant growth monitoring, and agronomic training, bridging the gap between real-world agricultural practices and advanced digital twin systems. 

Integrating editable NeRF and 3DGS frameworks, such as EditingNeRF \citep{liu2021editing}, NeRF-Editing \citep{yuan2022nerf}, Gaussian Grouping \citep{ye2024gaussian}, and SC-GS \citep{huang2024sc}, represents an initial step toward dynamic, editable NeRF and 3DGS rendering, allowing greater possibility in modifying plant structures and visualizing different growth stages. Future research should focus on developing interactive NeRF/3DGS models that support real-time editing, object-level segmentation, and predictive growth simulations, enhancing accuracy, scalability, and usability in virtual plant phenotyping while improving user control, visualization fidelity, and immersive agricultural analysis.

\subsubsection{Downstream applications}
NeRF and 3DGS have garnered increasing attention from the research community, leading to their application in various downstream tasks across agriculture, plant phenotyping, and environmental monitoring, as summarized in Table~\ref{tab:apps}. These techniques have been employed in high-throughput plant trait analysis, precision agriculture, and 3D scene understanding, demonstrating their versatility in automated plant monitoring and phenotypic trait extraction. Recent advancements have focused on integrating NeRF and 3DGS with state-of-the-art computer vision models to improve segmentation accuracy, reconstruction fidelity, and real-time processing. For instance, the Segment Anything Model (SAM) \citep{kirillov2023segment} has been combined with 3DGS for automated plant trait analysis, enabling precise organ-level segmentation in complex agricultural environments. 
To further enhance NeRF and 3DGS-based models, future research should focus on integrating cutting-edge deep learning and computer vision techniques for downstream applications, such as: automated fruit detection and yield estimation, crop disease monitoring, digital twin systems for smart farming.


\section{Summary}
\label{sec5}
Efficient and effective plant phenotyping methods are crucial for deepening our understanding of plant traits and their environmental interactions, paving the way for advancements in precision agriculture and crop improvement. 3D reconstruction technologies, including classical approaches, NeRF, and 3DGS, offer innovative tools to capture detailed plant morphology and enable automated, high-throughput phenotyping. However, these methods face distinct challenges that limit their widespread adoption. 
In this review, we explored recent developments in 3D reconstruction techniques, their applications in plant phenotyping, and the obstacles encountered in their implementation. 
We also examined potential solutions, such as optimizing classical methods for scalability, reducing NeRF’s computational burden, and leveraging 3DGS for enhanced efficiency. 
This study underscores the transformative potential of these 3D reconstruction approaches in agricultural research and highlights the need for ongoing innovation to fully realize their benefits in plant phenotyping.

\section*{Appendix}
In this analysis, the standard PRISMA (\underline{P}referred \underline{R}eporting \underline{I}tems for \underline{S}ystematic \underline{R}eviews and \underline{M}eta-\underline{A}nalysis) approach \citep{moher2009preferred} is utilized to methodically and exhaustively compile related studies. 
This technique follows established guidelines for systematic reviews, incorporating clearly outlined inclusion and exclusion standards \citep{li2023label, chen2025integrating}. 
To guarantee a comprehensive evaluation, relevant works were sourced from meticulously chosen databases. At the outset, prominent scientific platforms such as Web of Science, ScienceDirect, Springer, and Elsevier were used to identify key topics. 
Later, the search was expanded to include other resources like popular academic search tools (e.g., Google Scholar, IEEE Xplore) and freely accessible repositories such as ArXiv\footnote{ArXiv: \url{https://arxiv.org/}} to include cutting-edge developments. This thorough strategy is critical for documenting the latest 3D reconstruction methods and their use in plant phenotyping research.

Once the databases for gathering literature are determined, an \textit{inclusion} criterion is implemented to narrow down the article selection within the designated databases and search platforms. The procedure starts with a keyword-driven query, combining terms from two distinct fields. The initial group of keywords relates to 3D reconstruction, featuring expressions such as point cloud'', neural radiance fields'', and 3D modeling''. The second group centers on plant phenotyping, incorporating phrases like plant phenotyping'', precision farming'', crops'', and intelligent agriculture''. Logical operators, including ``AND'' and ``OR'', are used to optimize search precision and achieve broad coverage. Furthermore, during this \textit{inclusion} phase, the references and citations within the initially chosen articles are examined to widen the range of collected studies.

To select the most relevant studies for this review, a set of \textit{exclusion} criteria is employed to filter the articles gathered in the initial search. The first filter focuses on the publication timeline. Priority is given to works released within the past eight years (2018.01–2025.02) to maintain currency and capture the latest advancements, given that the use of sophisticated reconstruction techniques in plant phenotyping is a relatively new and fast-developing area. That said, earlier works with substantial citation numbers are retained due to their enduring influence in the domain. The second filter addresses article categories. Emphasis is placed on research papers featured in top-tier journals and conferences, while less detailed outputs like reports, meeting summaries, and similar publications are excluded due to their typically limited technical depth. Duplicate articles collected from various databases and search platforms are also eliminated. Following the application of these standards, 4 studies on NeRF (9 papers) and 3GS (3 Papers) in plant phenotyping were selected. Notably, only xx highly representative papers on traditional methods are included, as their exploration is already well-covered in existing review articles (refer to Table~\ref{tab:reviews}).


\section*{Acknowledgment}
The authors used large language models, such as ChatGPT, for grammar checking and language polishing; however, the original content was created by the authors, and the final version was reviewed and edited by them to ensure accuracy and clarity.

\typeout{}
\bibliography{ref}
\end{document}